\documentclass[12pt, draftclsnofoot, onecolumn]{IEEEtran}

\usepackage{config}
\usepackage{array}
\newcolumntype{P}[1]{>{\centering\arraybackslash}p{#1}}
\newcolumntype{M}[1]{>{\centering\arraybackslash}m{#1}}

\author{\IEEEauthorblockN{Abdoulaye Tall\IEEEauthorrefmark{1}, Zwi Altman\IEEEauthorrefmark{1}
and Eitan Altman\IEEEauthorrefmark{2}} \\ \IEEEauthorblockA{\IEEEauthorrefmark{1}
Orange Labs 38/40 rue du General Leclerc,92794 Issy-les-Moulineaux
\\Email: \{abdoulaye.tall,zwi.altman\}@orange.com}\\
\IEEEauthorblockA{\IEEEauthorrefmark{2}INRIA Sophia Antipolis, 06902
Sophia Antipolis, France, Email:eitan.altman@sophia.inria.fr}}

\title{Multilevel beamforming for high data rate communication in 5G networks}

\begin{document}
\maketitle
\begin{abstract}
    Large antenna arrays can be used to generate highly focused
    beams that support very high data rates and reduced energy
    consumption. However, optimal beam focusing requires large amount of feedback
    from the users in order to choose the best beam, especially in \ac{FDD} mode.
    This paper develops a methodology for designing a multilevel
    codebook of beams in an environment with low number of multipaths. The antenna design
    supporting the focused beams is formulated as an optimization
    problem. A multilevel codebook of beams is constructed according
    to the coverage requirements. An iterative beam scheduling is proposed that searches through the codebook
    to select the best beam for a given user. The methodology is
    applied to a mass event and to a rural scenario, both analyzed using an event-based network simulator.
    Very significant gains are obtained for both scenarios.
    It is shown that the more dominant the \ac{LoS}
    component, the higher the gain achieved by the multilevel
    beamforming.


\end{abstract}

\section{Introduction}

Different solutions and technologies have been developed for
focusing transmitted beams on users' receivers, in order to achieve
high bit rate communication or tracking capabilities. The
technological challenge is twofold: design antenna systems that
support narrow beams for large coverage area, and allow to locate
users with tight delay constraints for either scheduling or target
tracking. The considerable regain of interest in focusing antennas
is related to 5G systems which are in the process of definition in
different research frameworks and fora. It is expected that 5G
spectrum evolutions will benefit from higher frequency bands
\cite{METIS_IEEE_Magazine} making it possible to reduce the size of
large antenna arrays.

Different techniques have been developed to focus energy close to
the receiver. In an environment rich of multipaths which is
typically encountered in dense urban or indoor environment, \ac{TR},
also known as the Transmit Matched Filter, is receiving increasing
interest, particularly for the internet of things
\cite{chen2014InternetOfThings}. \ac{TR} is a pre-filtering
technique for \ac{MIMO} wireless systems
\cite{TimeReveral2004Paulraj}. The channel state information at
several transmit antennas is used to precode transmitted symbols
with the time reversed version of their respective channel impulse
responses. These "time reversed" waves propagate in the channel,
resulting in power focusing in space and time at the receiver.
\ac{TR} can be efficiently used for low complexity receivers such as
those expected in future connected objects \cite{DThuy2013dumb}, and
even for high mobility scenarios \cite{DThuy2013fastVehicles}. The
drawback of this technique is that it requires \ac{TDD} and full
\ac{CSI} at the transmitter side.

The problem of focusing the beam on moving targets has been addressed for decades
in the context of radar applications, and more specifically, in
millimeter wave phase array technology. Recently, the use of
millimeter adaptive antenna array, operating at the 70/80 GHz bands,
has been proposed for serving a high data rate moving user
\cite{SamsungPaper2010hybrid}. The antenna array consists of analog
sub-arrays behind a digital beamformer that estimates the \ac{AoA}
for the moving user.

Multilevel beamforming has been proposed in \cite{multileveBF2011backhaul} for millimeter wave backhaul
serving urban pico-cells. To ensure the link quality, an efficient search algorithm is performed to maintain
the alignment of the transmit and receive beams. The search is performed on a multilevel codebook thus reducing
considerably the number of beam couples to be tested. The channel gain matrix is not required, making this
technology cheaper and suitable for \ac{FDD} systems.

Recently, the concepts of \ac{VSC}
\cite{AnaMaria_virtual_smallcell_2015} and \ac{ViS}
\cite{tall_virtualsectorizationdesign_2015} have been proposed, with
the aim of creating a small cell or a sector using an  antenna array
located at the base station. To manage interference between the
macro cell and \ac{ViS}, a self-optimizing frequency splitting
algorithm has been proposed that dynamically shares the frequency
bandwidth between the two cells. The main technological challenge
for both \ac{VSC} and \ac{ViS} is how to optimally focus the beam at
the area where traffic is present. This issue is implicitly solved
by the multilevel beamforming proposed in this paper.

The present work adapts the concept of multilevel beamforming 
to the context of multi-user communication. In terms of
implementation complexity, the proposed approach is particularly
attractive. It can be implemented in a \ac{FDD} system where the
beam selection only depends on the mobile feedback, namely on the
\ac{CQI}. Different scenarios can be envisaged for the multilevel
beamforming, such as cellular coverage (in the sense of service
provisioning) in a low level or multipath propagation environment or
the coverage of mass events and crowded area. Unlike the backhaul
problem described above, the multilevel beamforming in this paper is generated at
the transmitter only. Cell coverage should be guaranteed with low
level inter-cell interference. The projection of beams (particularly
high gain narrow beams) may strongly spread out and overshoot
neighboring cells. Hence the generation of the set of multi-level
beams (or codebook) should take into account geometrical parameters
of the cell (e.g. cell size, antenna height) and can be performed
off-line.

    The contributions of the paper can be summarized as follows:
    \begin{itemize}
    \item An antenna optimization framework for generating beams with reduced interference from side-lobes for large angular domain.
    \item A beamforming codebook design strategy which avoids overshooting at neighboring cells.
    \item A beam selection algorithm for efficient search through the multilevel beamforming codebook.
    \item Performance analysis of the multilevel beamforming for two \ac{LTE} use cases taking into account different channel and traffic models.
    \end{itemize}

    The rest of the paper is organized as follows. Section \ref{sec:antenna_design} presents the antenna array design
    and optimization for the multilevel beamforming. Section \ref{sec:multilevel_beamforming}
    describes the multilevel beam construction and their selection algorithm. The performance of the
    proposed multilevel beamforming for different channel models is analyzed in Section \ref{sec:results} for a mass event
    and a rural scenario using an event-based \ac{LTE} network simulator.
    Section \ref{sec:conclusion} concludes the paper.

\section{Antenna Array Design} \label{sec:antenna_design}

    This Section presents the guidelines for the antenna array design supporting multilevel beamforming.
    The antenna array modeling is first introduced and then the design of the beam diagrams is formulated
    as an optimization problem.

\subsection{Antenna model}
\begin{figure}[!ht]
\centering
\includegraphics[width=3.5in]{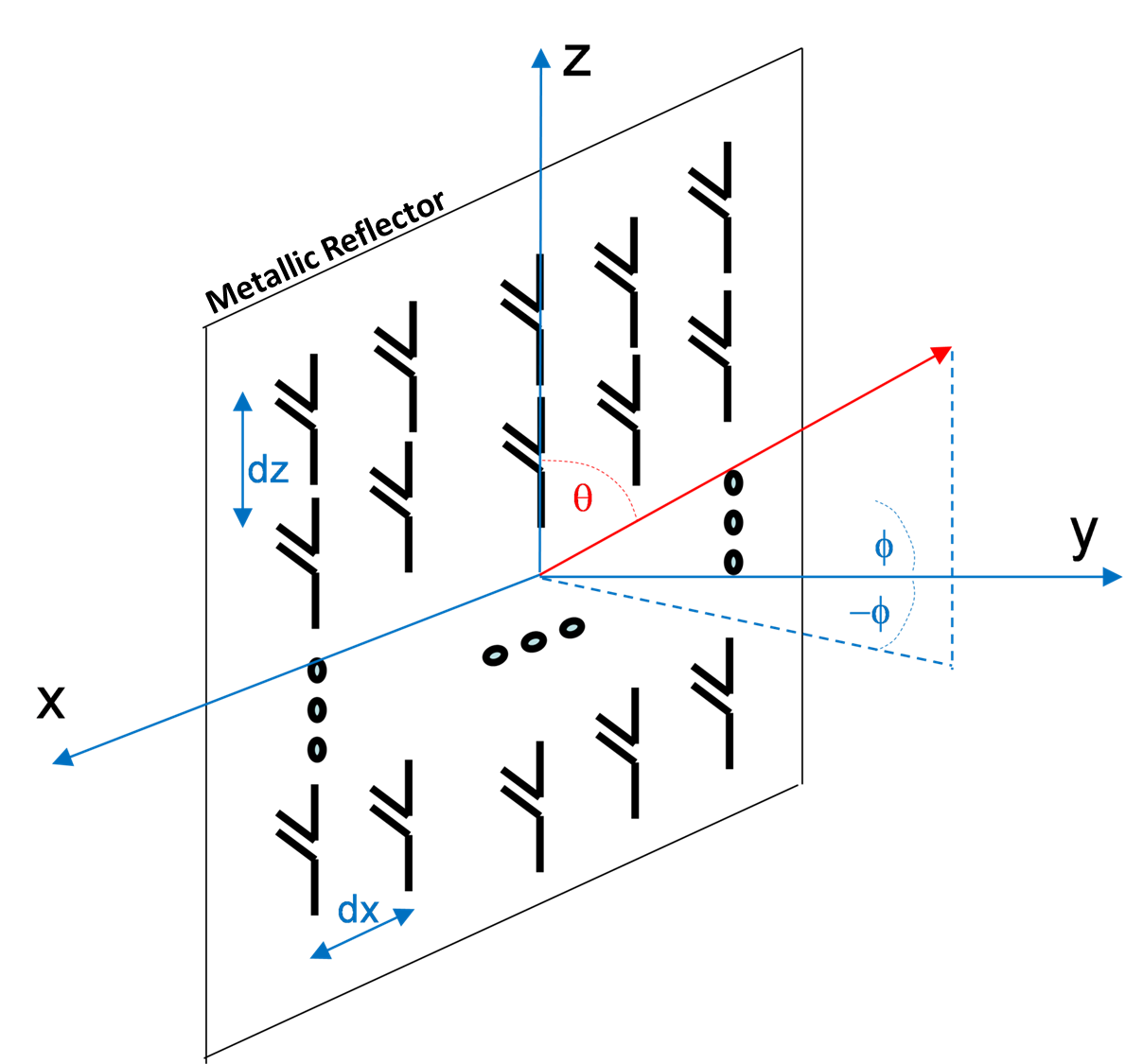}
\caption{Antenna array with dipole radiating elements.}
\label{fig:ant_array}
\end{figure}

    Consider a $N_x \times N_z$ (sub) antenna array of vertical dipoles, at a distance of  $\frac{\lambda}{4}$
    from a square metallic conductor, with $\lambda$ being the wavelength. The full antenna array generates the
    highest level (narrowest) beams, whereas lower level beams correspond to smaller (rectangular) sub arrays size (see Figure \ref{fig:beam_hierarchy}).
    If another type of radiating element is chosen, only its radiation pattern should be modified. To simplify the model,
    we approximate hereafter the reflector as an infinite \ac{PEC}. The $N_x$ and $N_z$  elements in each row and column are equally
    spaced with distances $d_x$ and $d_z$, respectively (Figure \ref{fig:ant_array}).

The direction of a beam is determined by the angle $(\theta_e,\phi_e)$ in the spherical coordinates
$\theta$ and $\phi$. The antenna gain for a given beam defined by $(\theta_e,\phi_e)$ in a given direction $(\theta, \phi)$ is written as
\begin{equation}
    G(\theta,\phi,\theta_e,\phi_e) = G_0 f(\theta,\phi,\theta_e,\phi_e )
\end{equation}
where $f$ is a normalized gain function and $G_0$ the maximum
gain in the $(\theta_e,\phi_e)$ direction. A separable excitation in
the $x$ and $z$ directions is assumed, resulting in the following
separable form of $f$:
    \begin{equation}
    \begin{split}
        f(\theta,\phi,\theta_e,\phi_e ) = | & AF_x^2 (\theta,\phi,\theta_e,\phi_e ) \cdot AF_y^2 (\theta,\phi) \\
        & \cdot AF_z^2 (\theta,\theta_e )| \cdot G_d (\theta)
    \end{split}
    \end{equation}

    $AF_x (\theta,\theta_e,\phi,\phi_e)$ and $AF_z(\theta,\theta_e)$ are the array factors in the $x$ and $z$ directions and are given by
    \begin{equation}
        AF_x (\theta,\theta_e,\phi,\phi_e )=\frac{1}{\sum_{m=1}^{N_x} w_m} \sum_{m=1}^{N_x} w_m \cdot a_m
    \end{equation}
and
    \begin{equation}
        AF_z (\theta,\theta_e )=\frac{1}{\sum_{n=1}^{N_z} v_n} \sum_{n=1}^{N_z} v_n \cdot b_n .
    \end{equation}

    $a_m$ and $b_n$ are complex amplitude contributions of the radiating element
    located at $(m-1) d_x$ and $(n-1) d_z$, respectively:
    \begin{equation}
        a_m = \exp\left(-j 2 \pi \frac{(m-1)d_x}{\lambda} (\sin\theta\sin\phi-\sin\theta_e\sin\phi_e)\right),
    \end{equation}
    \begin{equation}
        b_n=\exp \left(-j 2 \pi \frac{(n-1)d_z}{\lambda} (\cos\theta-\cos\theta_e) \right).
    \end{equation}

    The weights $w_m$ and $v_n$ for radiating elements in the $m$-th row and $n$-th columns define a Gaussian tapering
    function used to control the sidelobe level of the gain pattern
    \begin{equation}
        w_m = \exp{\left(-\left(\frac{(m-1) L_x - \frac{L_x}{2}}{\sigma_x} \right)^2 \right)},
    \end{equation}
    \begin{equation}
        v_n = \exp{\left(-\left(\frac{(n-1) L_z - \frac{L_z}{2}}{\sigma_z} \right)^2 \right)},
    \end{equation}
    where $L_x$ and $L_z$ are the array size in the $x$ and $z$ directions respectively, with $L_x = (N_x-1) d_x$ and $L_z=(N_z-1) d_z$.
    The values for $\sigma_s, s=x,z$, are defined by fixing the ratio between the extreme and center dipole amplitudes respectively to a given value of $\alpha_s$:
    \begin{equation}
        \sigma_s^2=- \left( \frac{L_s}{2} \right)  \frac{1}{\log(\alpha_s )}    ;  s=x,z
    \end{equation}

    The impact of the \ac{PEC} can be modeled by replacing it with the images of the radiating elements it
    creates. The term $AF_y(\theta,\phi)$ takes into account the images and is written as:
    \begin{equation}
        AF_y (\theta,\phi) = \sin\left(\frac{\pi}{2} \sin\theta \cos\phi \right)
    \end{equation}

    The normalized gain pattern of the dipoles, $G_d(\theta)$, is approximated as
    \begin{equation}
        G_d(\theta) = \sin^3\theta .
    \end{equation}

    The term $G_0$ is obtained from the power conservation equation:
    \begin{equation}
        G_0 = \frac{4\pi}{\int_{-\frac{\pi}{2}}^{\frac{\pi}{2}} \int_0^\pi f(\theta,\phi) \sin\theta \, d\theta d\phi}.
    \end{equation}

    A beam is defined by the (rectangular) sub-array size, and the couple $(\theta_e,\phi_e)$ defines its direction.
    The generation of the beams is discussed in Section \ref{sec:multilevel_beamforming}. The antenna modeling for all the
    beams remains unchanged.

\subsection{Antenna pattern optimization}

    The (sub) antenna array design constitutes an optimization problem with two objectives: maximizing the antenna gain
    (or conversely, minimizing the width of the main lobe) and minimizing the side-lobes' level, as a function of the
    parameters $d_s$ and $\alpha_s$, $s=x,z$. The problem is written as a constrained optimization problem:
    \begin{subequations} \label{eq:ant_opt}
    \begin{align}
    \underset{d_x,d_z,\alpha_x,\alpha_z}{\text{maximize}} \, & G_0(N_{x,\max},N_{z,max},d_x,d_z,\theta_e,\phi_e) \\
         \text{s.t. } & \\
        & \max_{N_x,N_z}\{\emph{SL}(N_x,N_z,d_x,d_z,\theta_e,\phi_e)\} \leq  Th; \label{eq:sl_th} \\
        & \,\, N_{s,\min} \leq N_s \leq N_{s,\max}; \, s=x,z; \\
        & \,\, 0 < \alpha_s \leq 1 ; \, s=x,z; \\
        & \,\, 0 < d_s \leq \frac{\lambda}{2} ; \, s=x,z ;\\
        & \,\, \theta_\text{min} \leq \theta_e \leq \theta_\text{max} ; \\
        & \,\, - \phi_\text{max} \leq \phi_e \leq \phi_\text{max};
    \end{align}
    \end{subequations}
where $N_{s,\min}$ and $N_{s,\max}$ are respectively the minimum and
maximum number of antenna elements in the $s$ direction,
$\theta_\text{min}, \theta_\text{max}, \phi_\text{min},
\phi_\text{max}$ are respectively the minimum and maximum electrical
elevation and azimuth angles of the antenna array. The constraint
\eqref{eq:sl_th} reads: the maximum side-lobe level for the whole
range of sub array size should be below a predefined threshold $Th$.

    It is noted that for small elevation electrical tilt values, the projection of the beams on the plane is likely to spread out.
    Special care should be taken when setting the $\theta_\text{min}, \phi_\text{min}, \phi_\text{max}$ angles in order to avoid overshooting
    on neighboring cells. These angles will depend on the geometrical characteristics of the cell (original coverage area of the considered \ac{BS} before deploying the antenna array), such as the cell shape, size, and antenna height.
    One can consider optimizing the antenna for a wide range of elevation and azimuth angles, and then, according to the cell geometry, construct
    a codebook of beams for a desired angular range. Furthermore, a database with a set of codebooks can be pre-optimized for a set of cell geometries and
    then, according to the specific cell deployment, the most suitable codebook can be selected.


\section{Multilevel beamforming} \label{sec:multilevel_beamforming}

\subsection{Beams structure}

\begin{figure}[!ht]
\centering
\includegraphics[width=2in]{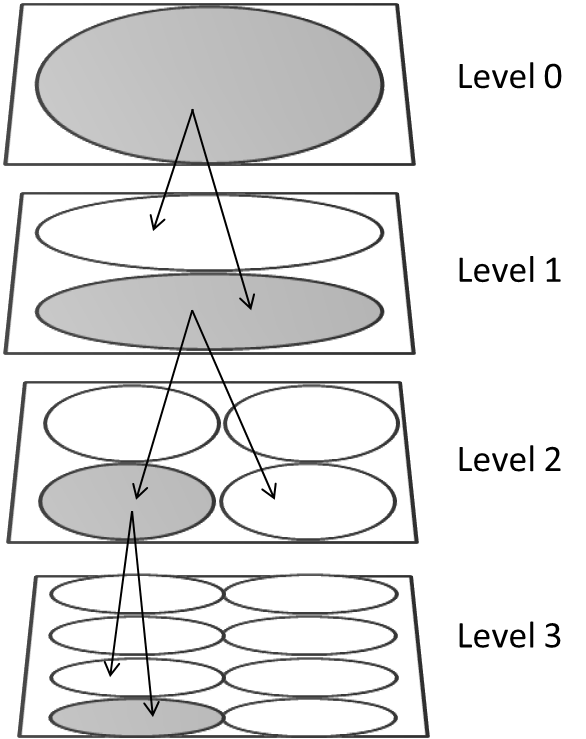}
\caption{Example of beam hierarchy}
\label{fig:beam_hierarchy}
\end{figure}

    Consider a multilevel codebook as shown in Figure \ref{fig:beam_hierarchy}, with $L$ levels and $J_l$ beams
    at the level \emph{l}, $l=0,...,L$. The $j$th beam at level $l$ is written as $B_{l,j}
    (\theta,\theta_{l,j},\phi,\phi_{l,j})$, $j=1,...,J_l$, and for brevity of notation - as $B_l(j)$.
    It is noted that the angles $(\theta_{l,j},\phi_{l,j})$ correspond to ($\theta_e,\phi_e$) defined
    in section \ref{sec:antenna_design}.

    The beam of the first level, namely level $0$ in Figure \ref{fig:beam_hierarchy}, $B_0(1)$, covers the entire cell.
    Beams at level $l$ are generated by the same sub-array, i.e.
    with the same number of array elements, and differ from each
    other by the angles $(\theta_{l,j},\phi_{l,j})$. Denote by $\hat{C}$
    a \emph{coverage operator} which receives as argument a beam, and
    outputs its coverage area (often denoted as the best server area),
    where the beam provides the strongest signal with respect to other
    cell or beam coverage. By construction we assume that for a
    given level $l$, the beams' coverage do not overlap:
    \begin{equation} \label{eq:coverage_overlap}
        \hat{C}(B_l(i)) \wedge \hat{C}(B_l(j)) = \emptyset \, , \forall i \neq j.
    \end{equation}

    The multilevel structure of the beams in the codebook means that a given beam $B_l(j)$ at level $l$ where $l < L$
    has two children beams $B_{l+1}(2j-1)$ and $B_{l+1}(2j)$ with
    \begin{equation} \label{eq:coverage_inclusion}
     \hat{C}(B_{l+1}(2j-1)) \cup \hat{C}(B_{l+1}(2j)) \subset \hat{C}(B_{l}(j))
    \end{equation}
    The beams at level $L$ are the narrowest that can be obtained given the $N_{x,max} \times N_{z,max}$ antenna array.

\subsection{Beam selection algorithm}

    The beam selection algorithm consists in finding the best beam available by navigating through the multilevel codebook.
    It starts with $B_0(1)$ which covers the entire cell and keeps track of the overall best beam up till now ($B^*$).
    Assuming the beam selection algorithm is at level $l$ with $l < L$, the best beam is updated as follows
    \begin{align} \label{eq:beam_select}
    B^* &= \argmax{B \in \{ B^*, B_{l+1}(j), j = (2j_{l+1}-1,2j_{l+1}) \}} \, S(B,u)
    \end{align}
    where $S(B,u)$ is the \ac{SINR} of user $u$ served by the beam $B$, and $j_l$ denotes
    index of the beam at level $l$.

    The algorithm stops when the best beam does not change in a
    given iteration, i.e. the parent beam provides better \ac{SINR}
    than the children beams, or the highest level of beams $L$ is
    reached. The complexity of such an algorithm is of the order of
    $log(N)$, $N$ being the total number of beams, hence
    convergence is obtained in a very small number of iterations.

    It is noted that condition \eqref{eq:coverage_inclusion} can be relaxed so that narrower beams
    can cover regions not covered by their parent beams. In this case the beam selection algorithm
    should continue until level $L$. The multilevel beam structure presented in this section is just
    one illustration of how a multilevel codebook structure can be designed. Other approaches can be adopted regarding
    the relation between parent and children beams in terms of coverage. The only requirement is to be
    able to easily navigate through the codebook in an iterative manner.

    The beam selection algorithm runs in parallel with the scheduling algorithm. A user is first
    scheduled based on its \ac{SINR} obtained with the level 0 beam.
    During the scheduling period, the user's best beam is updated based on the received feedback using equation \eqref{eq:beam_select}.
    If the scheduling period is long enough, the beam selection algorithm can converge. Otherwise the beam selection
    resumes at the next scheduling period based on the \ac{SINR} of the best beam tested so
    far. It is noted that other scheduling strategies can be considered.

\subsection{Implementation issues}
    The beam codebook can be precalculated for a given cell and stored in a database of a
    self-configuration server at the management plane. Upon deployment of the multilevel beamforming feature,
    the multilevel codebook is downloaded from the server to the
    \ac{BS}. As mentioned previously, the codebook selection can be
    based on geometrical characteristics of the cell.

    The antenna array can be dynamically configured using different approaches. The classical approach would be to feed
    each antenna element by a distinct amplifier which receives the appropriate input
    signal necessary to excite the selected beam from the codebook.
    More recently, the \emph{load modulated massive MIMO} approach has been reported
    \cite{modulated_MIMO} in which the base band input signal is used to adapt a load behind each antenna
    element which controls its complex input impedance. This approach which utilizes a single amplifier
    aims at further reducing the antenna size and cost, and could be a
    candidate technology for the multilayer beamforming (further studies are still necessary).

    The size of the antenna array depends on the number of antenna elements and the spacing
    between them. In Section \ref{sec:mass_event}, we use $N_{x,max} = 12$ and $N_{z,max} = 32$
    antenna elements with spacings of $d_x = 0.5\lambda$ and $d_z =
    0.7\lambda$. For the \ac{LTE} technology with a 2.6 GHz
    carrier, the antenna size is of 0.69 m $\times$ 2.58 m.
    Multilevel beamforming will be particularly attractive in 5G technology where higher frequency bands will be 		available, allowing moderate size of antenna array with a large number of radiating elements.

\section{Numerical results} \label{sec:results}

    We present in this section numerical results for the multilevel beamforming.
    Two scenarios are considered: a mass event type of scenario in an urban environment
    with a crowded open area, e.g. an esplanade, and a rural environment in which
    the users have a \ac{LoS} path component from the \ac{BS}. The multilevel beamforming is applied to one cell
    which is interfered by two rings of neighboring \acp{BS}.

    A \ac{LTE} event-based simulator coded in Matlab is used. Users (data sessions) arrive according to a Poisson process,
    download a file of exponential size with mean of 4 Mbits, and leave the network as soon as their download is complete. We focus on the case where there is no mobility.
    The channel coherence time of several milliseconds is assumed
    (which is typically the case for low mobility), and so the beam selection algorithm converges within this coherence time.
    Hence beam selection errors due to fast-fading are not considered.

    The main simulation parameters used for the two scenarios are summarized in Table \ref{tab:paramsg}.
    \begin{table}[!t]
    \small
    \renewcommand{\arraystretch}{1.3}
    \caption{Network and Traffic characteristics}
    \label{tab:paramsg}
    \centering
    \begin{tabular}{|M{4cm}|M{3cm}|}
    \hline
    \multicolumn{2}{|c|}{Network parameters} \\
    \hline
    Number of sectors with multilevel beamforming & 1 \\
    \hline
    Number of interfering macros & 2 rings, 20 sectors \\
    \hline
    Macro Cell layout & hexagonal trisector \\
    \hline
    Antenna height & 30 m \\
    \hline
    Bandwidth & 10MHz \\
    \hline
    Scheduling Type & Proportional Fair \\
    \hline
    \multicolumn{2}{|c|}{Channel characteristics} \\
    \hline
    Thermal noise & -174 dBm/Hz \\
    \hline
    Path loss ($d$ in km) & 128.1 + 37.6 $\log_{10}(d)$ dB \\
    \hline
    Shadowing & Log-normal (6dB) \\
    \hline
    \multicolumn{2}{|c|}{Traffic characteristics} \\
    \hline
    Service type & FTP \\
    \hline
    Average file size & 4 Mbits \\
    \hline
    \end{tabular}
    \end{table}

    \subsection{Mass event scenario} \label{sec:mass_event}

\begin{figure}[!ht]
    \centering
    \includegraphics[width=3in]{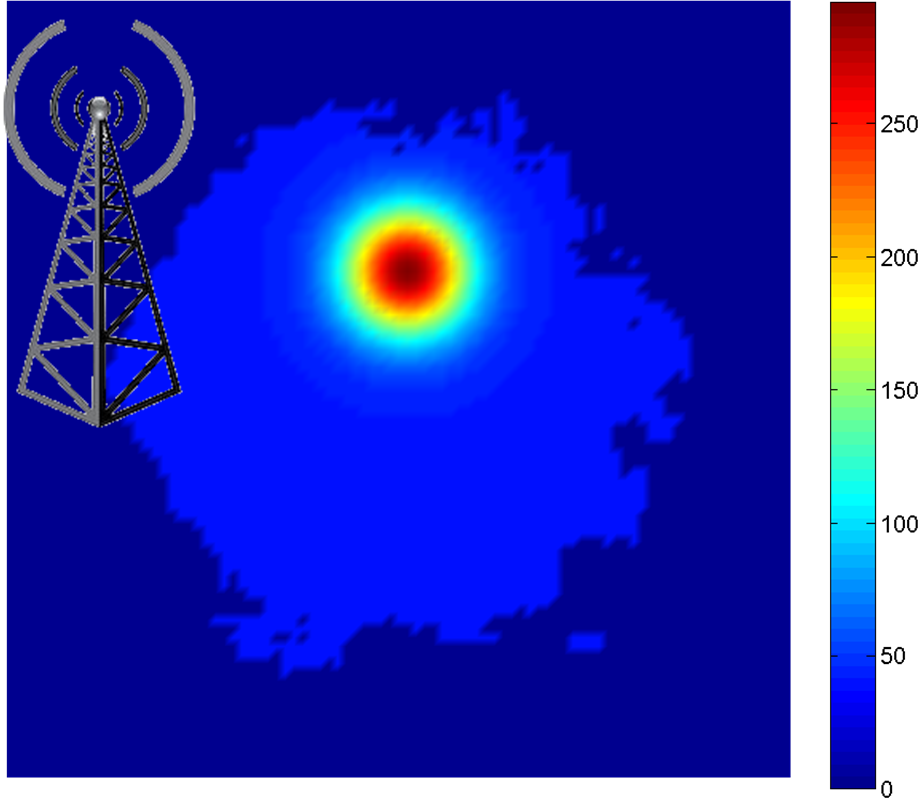}
    \caption{Traffic intensity map (in users/s/km$^2$).}
    \label{fig:traff_map}
    \end{figure}

    Consider a cell with a very large hotspot described by a spatial Gaussian traffic distribution which is superimposed
    on a uniform traffic over the whole cell area as shown in Figure \ref{fig:traff_map}. This distribution
    represents the traffic intensity map in arrivals per second per km$^2$. The hotspot can represent a crowd watching a live concert
    held on an esplanade for example.
    It is assumed that the users have a significant direct (\ac{LoS}) path with the \ac{BS}.
    It is recalled that in a rich scattering environment, other \ac{MIMO}
    techniques are more appropriate. In order to take into account the residual multipaths due to
    reflections on neighboring buildings, we use the Nakagami-m distribution for the fast fading.

    \subsubsection{Illustration of the multilevel beamforming}

\begin{figure}
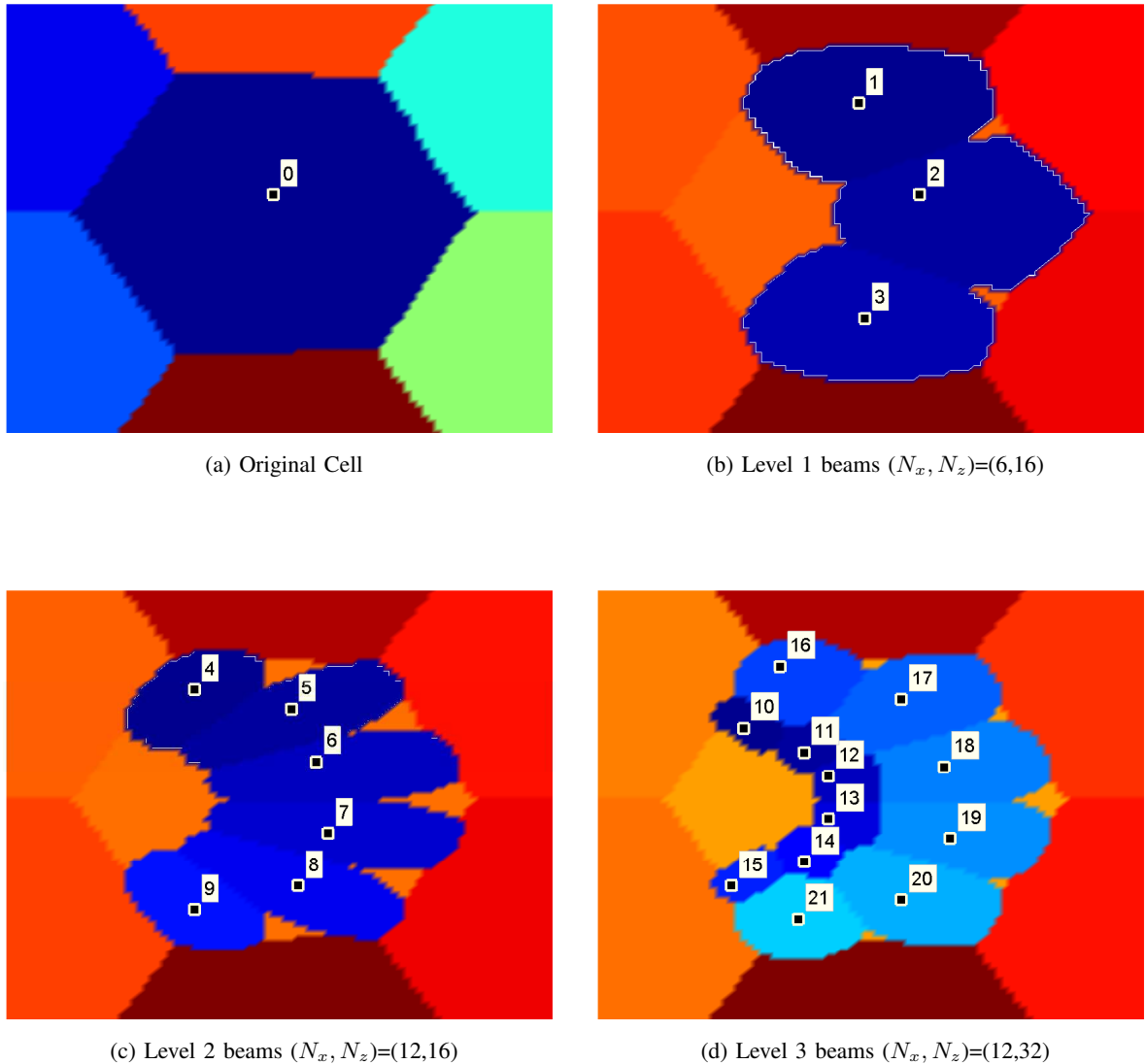

        \centering
        \subfloat[Original Cell]{\includegraphics[width=3in]{beam0}
        \label{fig:beam0u}}
        \hfil
        \subfloat[Level 1 beams ($N_x,N_z$)=(6,16)]{\includegraphics[width=3in]{beam1}
        \label{fig:beam1u}}
        \hfil
        \subfloat[Level 2 beams ($N_x,N_z$)=(12,16)]{\includegraphics[width=3in]{beam2}
        \label{fig:beam2u}}
        \hfil
        \subfloat[Level 3 beams ($N_x,N_z$)=(12,32)]{\includegraphics[width=3in]{beam3}
        \label{fig:beam3u}}
        \caption{Multilevel beam coverage maps for the mass event scenario}\label{fig:beams_u}
\end{figure}

\begin{figure}
        \centering
        \subfloat[Level 0 beam (\ac{SINR} = 7.75dB)]{\includegraphics[width=3in]{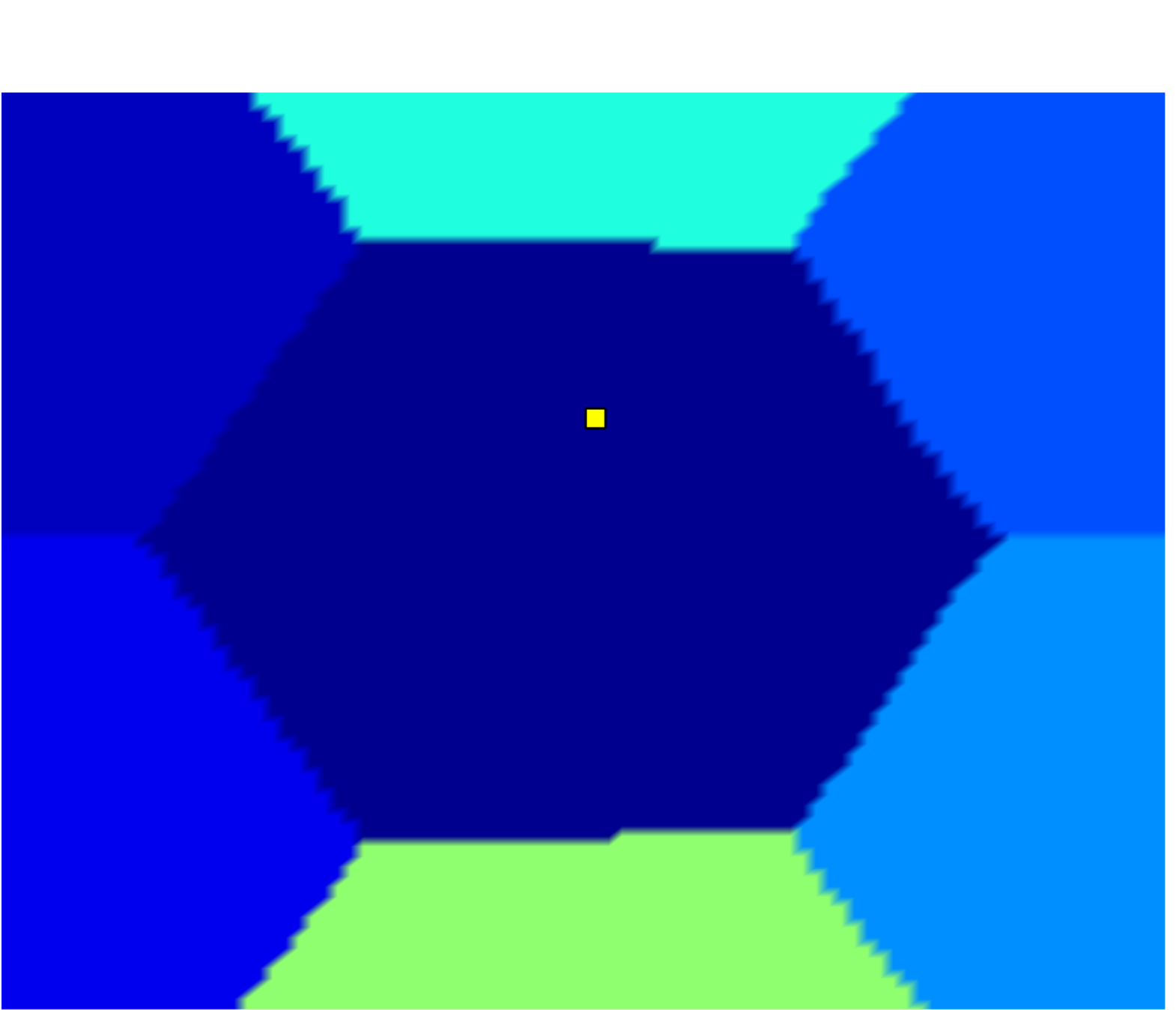}
        \label{fig:zoom0}}
        \hfil
        \subfloat[Level 1 beam (\ac{SINR} = 12.52dB)]{\includegraphics[width=3in]{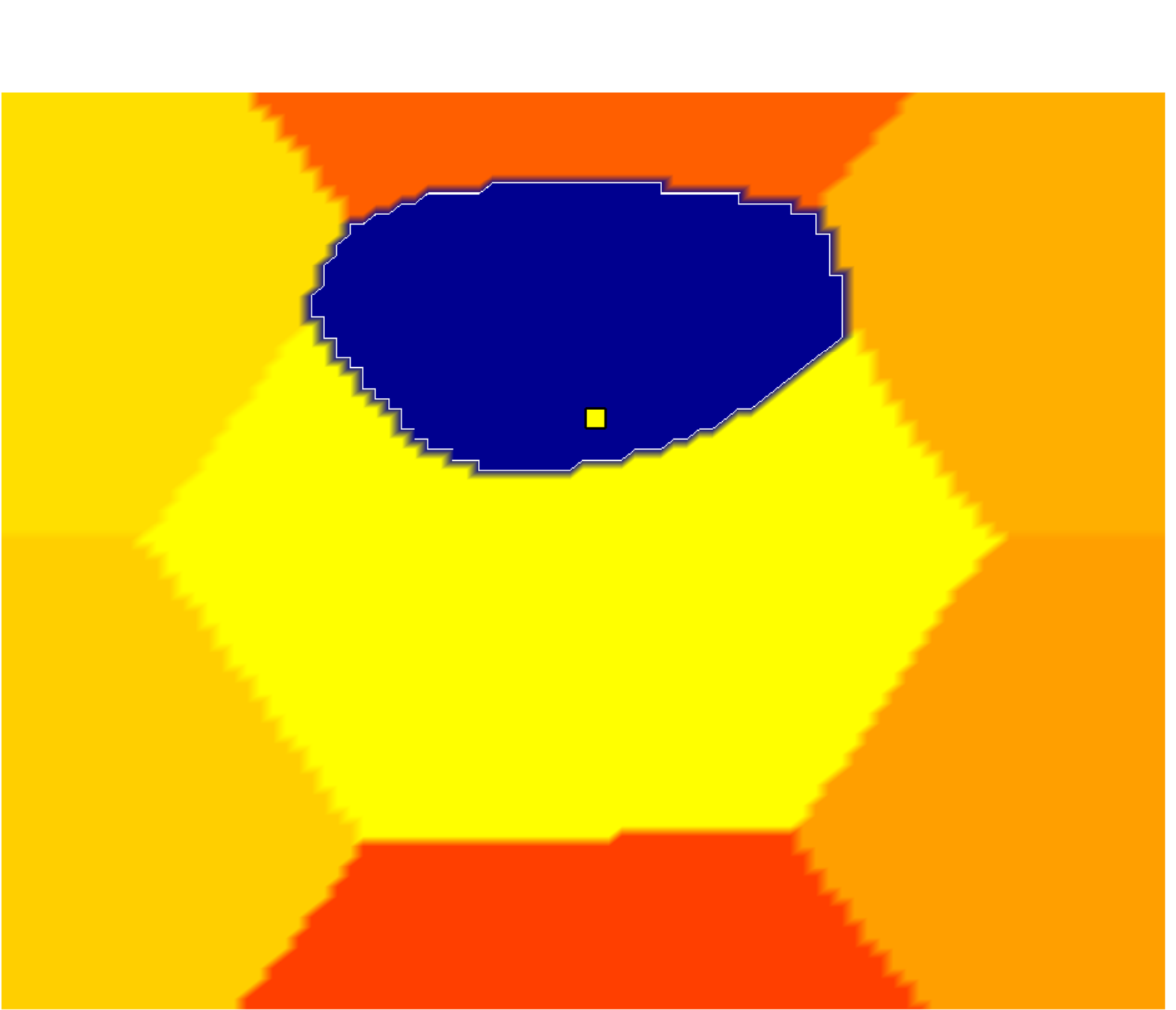}
        \label{fig:zoom1}}
        \hfil
        \subfloat[Level 2 beam (\ac{SINR} = 14.24dB)]{\includegraphics[width=3in]{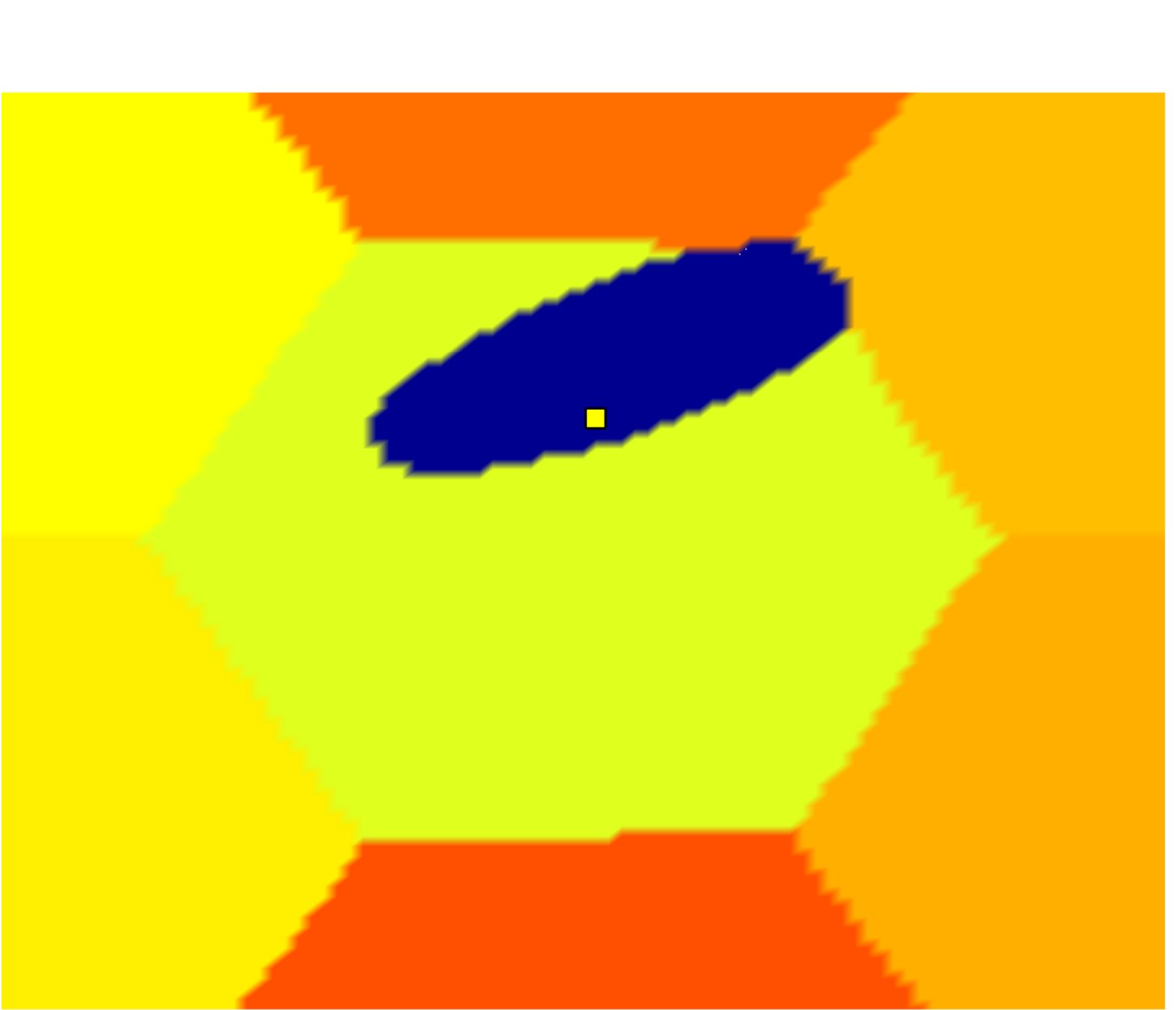}
        \label{fig:zoom2}}
        \hfil
        \subfloat[Level 3 beam (\ac{SINR} = 17.38dB)]{\includegraphics[width=3in]{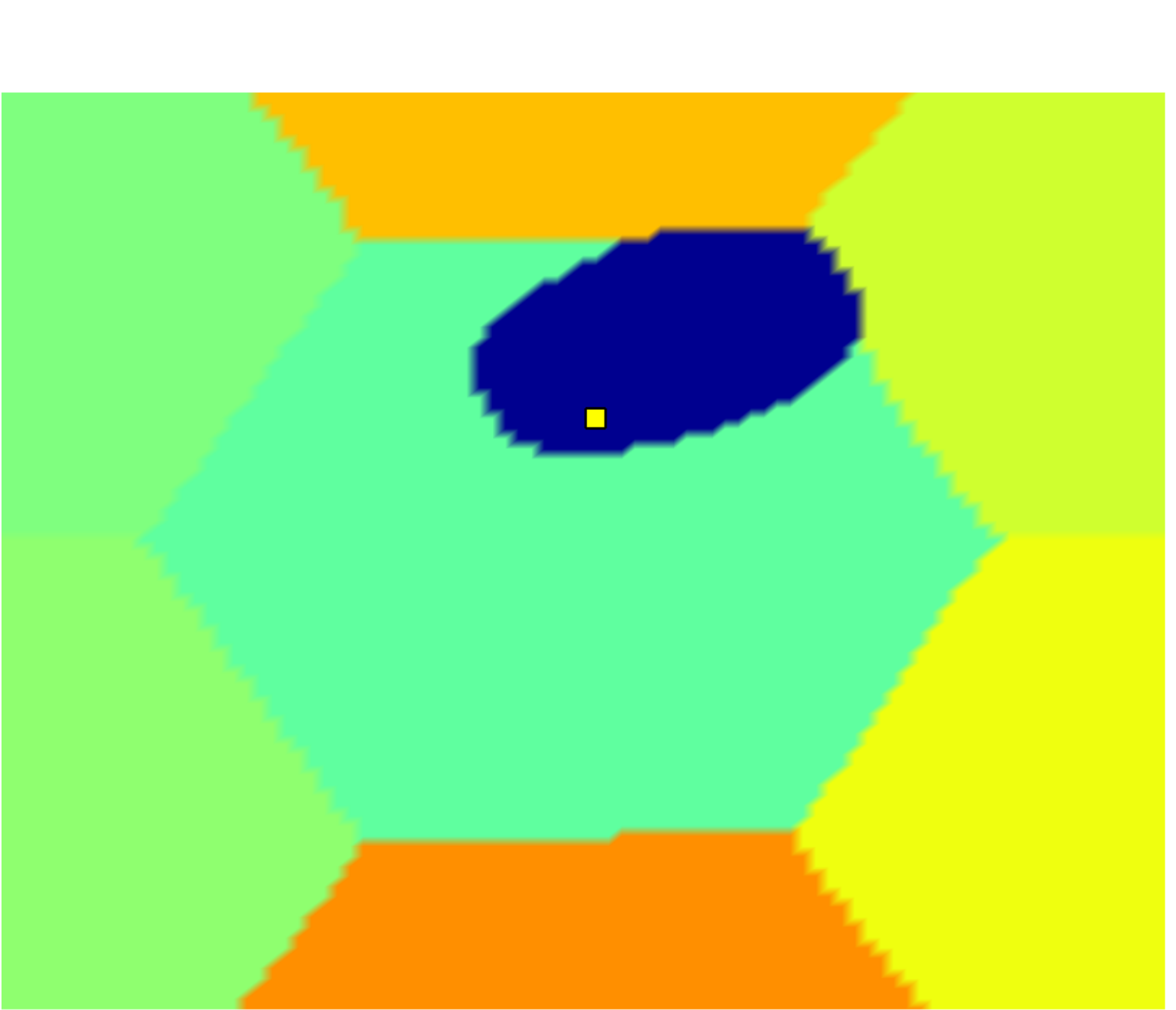}
        \label{fig:zoom3}}
        \caption{Successive narrowing of the beam for a given user}\label{fig:zooms}
\end{figure}

\begin{figure}
        \centering
        \subfloat[Level 1 beam diagrams]{\includegraphics[width=3in]{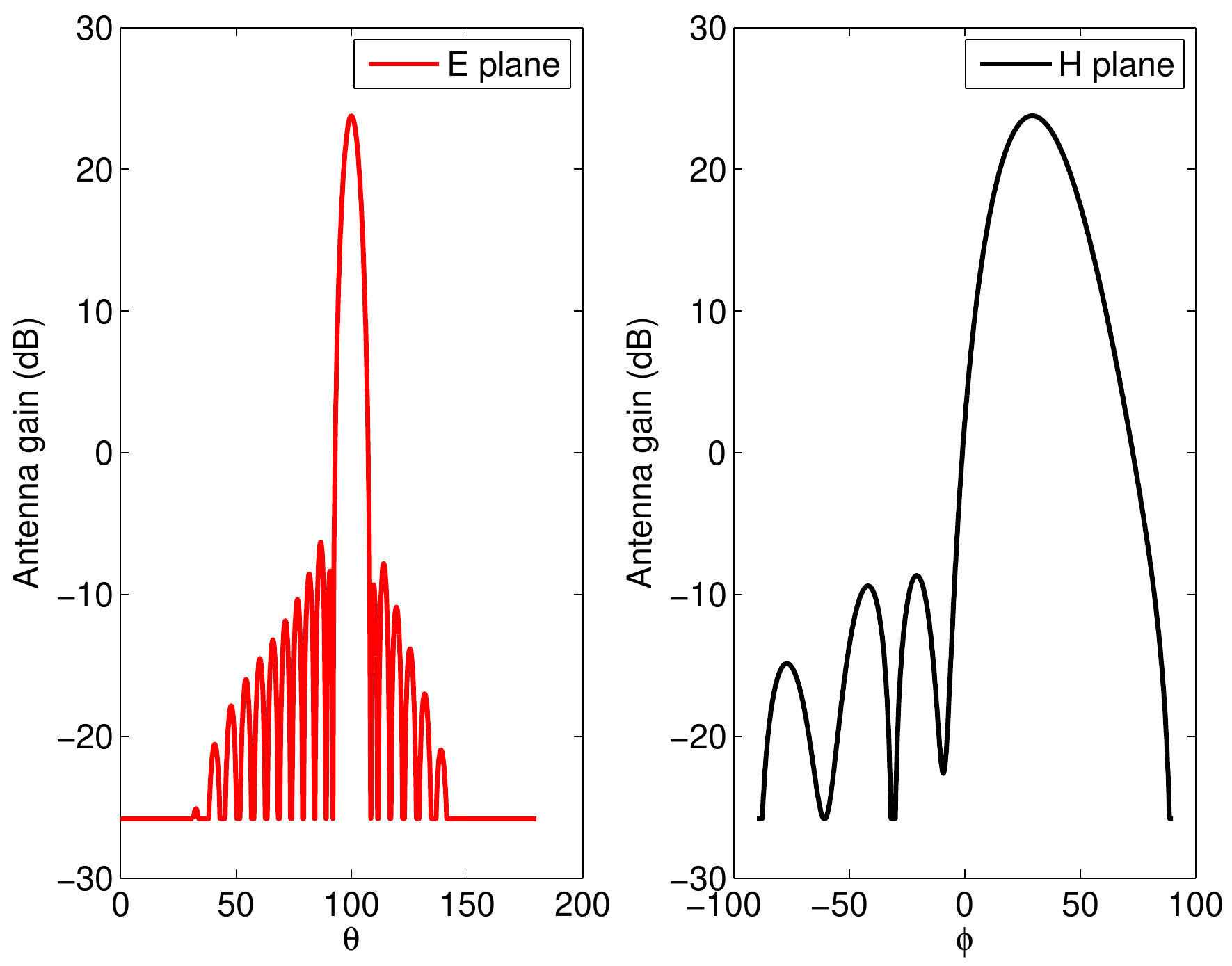}
        \label{fig:ant_diag1}}
        \hfil
        \subfloat[Level 2 beam diagrams]{\includegraphics[width=3in]{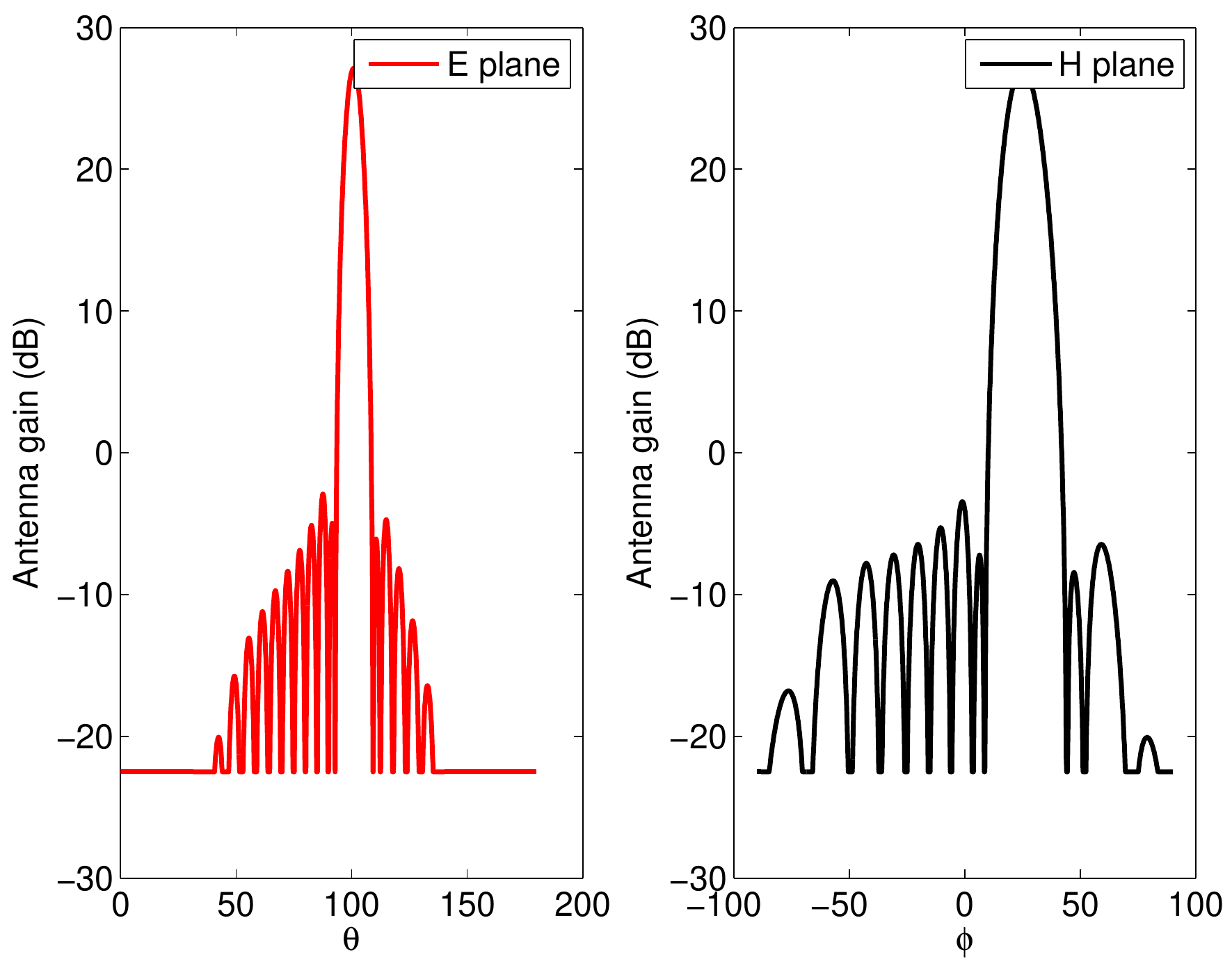}
        \label{fig:ant_diag2}}
        \hfil
        \subfloat[Level 3 beam diagrams]{\includegraphics[width=3in]{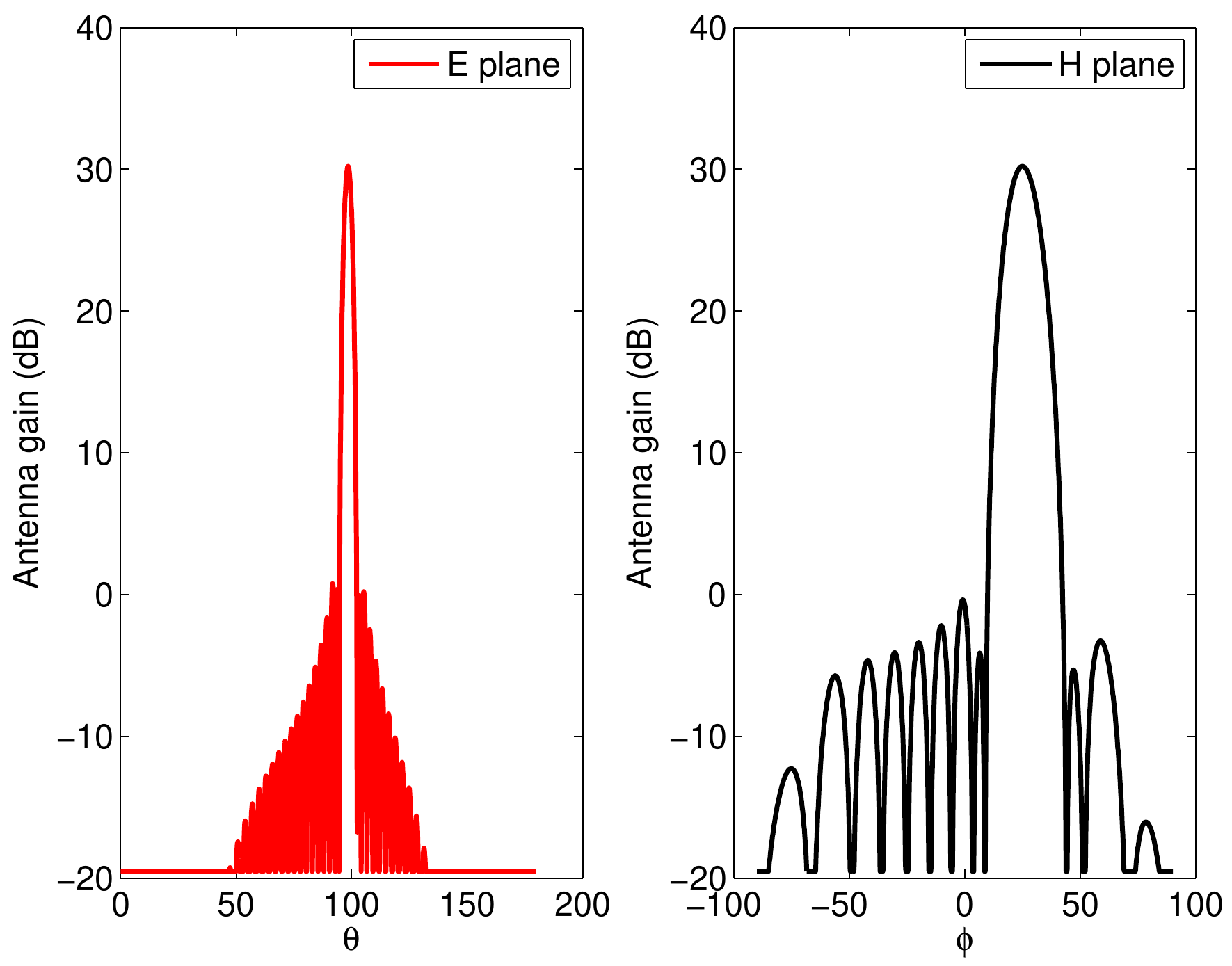}
        \label{fig:ant_diag3}}
        \caption{Antenna diagrams for a given user's best beam in each level.}\label{fig:ant_diags}
\end{figure}

    Figures \ref{fig:beams_u}, \ref{fig:zooms} and \ref{fig:ant_diags}
    represent the coverage maps of the beams in each level, the best beam chosen at each level for a user located at the center of the hotspot area
    (yellow square in Figure \ref{fig:zooms}) and the corresponding antenna diagrams, respectively, as described below.

    The multilevel beam structure presented in Section \ref{sec:multilevel_beamforming} is illustrated in Figure \ref{fig:beams_u}.
    Here, the condition \eqref{eq:coverage_overlap} is met by definition
    of the coverage areas. However, condition \eqref{eq:coverage_inclusion}
    is relaxed in order to allow narrower beams (level 3 in Figure \ref{fig:beam3u}) to cover blank spaces left by wider beams (level 2
    in Figure \ref{fig:beam2u}).

    Figure \ref{fig:zooms} presents the beam selection algorithm performed according to \eqref{eq:beam_select} for a given user.
    The \ac{SINR} of the selected user gradually increases from 7.75dB to 17.38dB at three iterations, i.e. by a factor of 9.
    It is noted that the \ac{SINR} gains are expected to be even higher for cell edge users.

    The antenna diagrams corresponding to the selected beams in each level in Figure \ref{fig:zooms} are presented in Figure \ref{fig:ant_diags}.
    These diagrams were designed according to the optimization problem \eqref{eq:ant_opt}, with the side-lobe level constraint
    \eqref{eq:sl_th} of $Th = 30dB$. One can see that the beam width of the main lobe gets narrower  in elevation or azimuth plane and the maximum gain increases (from 23.76dB to 30.2dB) with the beam level. The antenna diagram for level 0 which correspond to the full cell coverage is omitted.

\subsubsection{Performance results} \label{sec:urban_perf}
    We next evaluate the performance of the multilevel beamforming for the mass event urban scenario.
    We use the Nakagami-m distribution which models fast-fading in environments with strong \ac{LoS} component
    and many weaker reflection components. The \emph{shape parameter} m dictates
    the contribution of the \ac{LoS} component in the overall signal. For m $=1$, there is no \ac{LoS} component
    and the fast fading reduces to a Rayleigh distribution. As m grows to infinity, the \ac{LoS} component gradually becomes
    preponderant. We consider various Nakagami-m fading scenarios with m = 2, 5 and 10,
    and the no-fading case (corresponding to m $=+\infty$). We do not consider the m $=1$ case
    due to the open environment considered in the scenario. The simulation parameters for the scenario are summarized in Table \ref{tab:paramsu}.
\begin{table}[!t]
\small
\renewcommand{\arraystretch}{1.3}
\caption{Network and traffic characteristics for the mass event scenario} \label{tab:paramsu}
\centering
\begin{tabular}{|M{4cm}|M{4cm}|}
\hline
Intersite distance & 500 m \\
\hline
Nakagami-m shape parameter & 2, 5 or 10 \\
\hline
Traffic spatial distribution & Gaussian Hotspot + Uniform (see Figure \ref{fig:traff_map}) \\
\hline
\end{tabular}
\end{table}

\begin{figure}[!ht]
\centering
\includegraphics[width=3.5in]{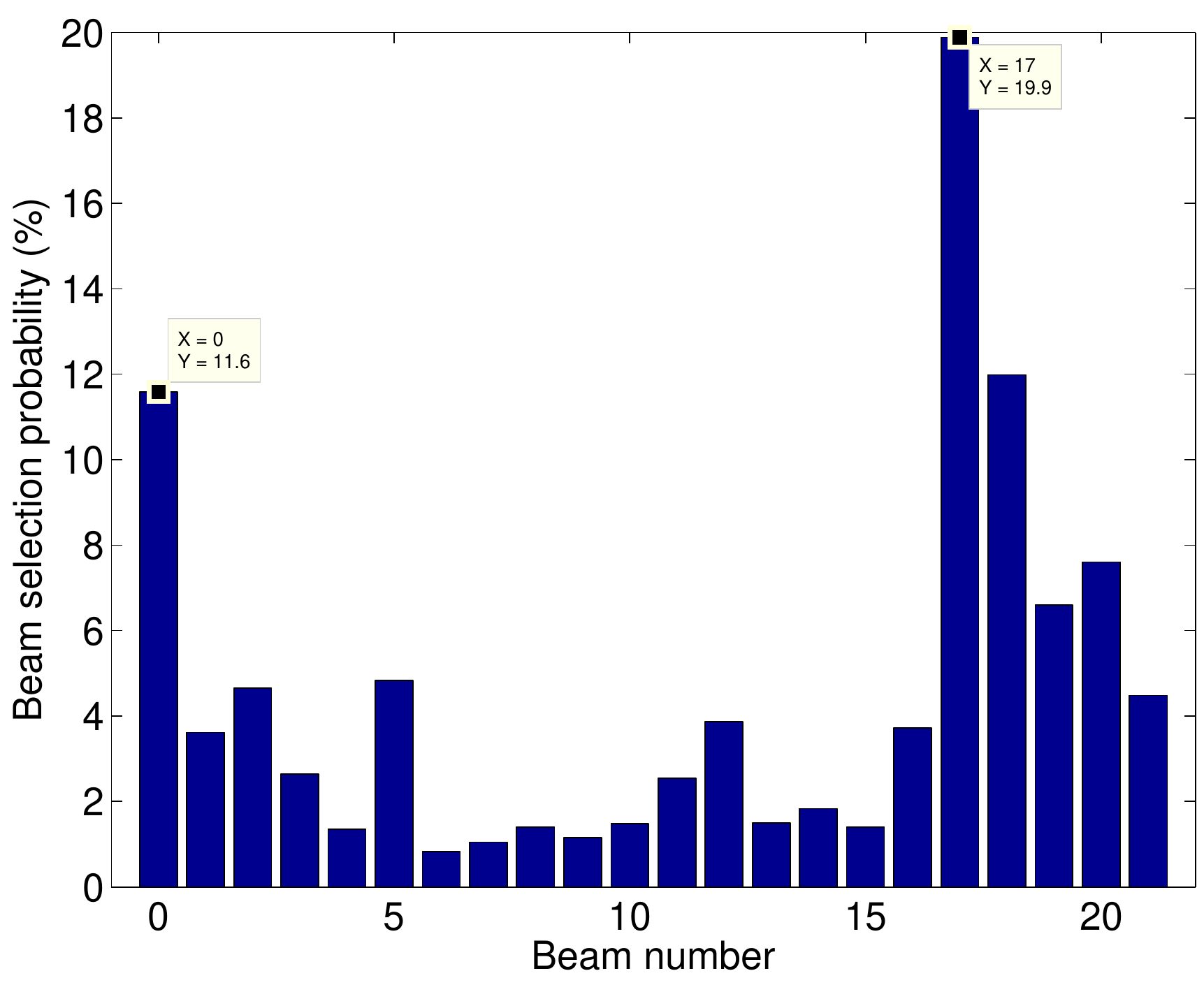}
\caption{Histogram of selected beams throughout the simulation: X is the beam number (see Figure \ref{fig:beams_u}) and Y - its selection probability.}
\label{fig:sel_beam_prob}
\end{figure}

    Figure \ref{fig:sel_beam_prob} presents the frequencies of selected beams throughout the simulation.
    As expected, beam 17 which covers most of the hotspot region (see Figures \ref{fig:traff_map} and \ref{fig:beam3u})
    is the most frequently selected. So the beam selection algorithm successfully locates the traffic in the direction of the hotspot and
    adjusts the beam width without any prior knowledge of the hotspot location and size.

    Table \ref{tab:kpisu} presents the \ac{MUT}, the \ac{CET} and the \ac{PC} obtained for the various shape parameters of
    the Nakagami-m fading distribution, with and without (denoted respectively as 'w.' and 'wo.' in Table \ref{tab:kpisu}) the multilevel beamforming.
    For example, in Table \ref{tab:kpisu}, '2 wo.' means m=2 without multilevel beamforming. The \ac{PC} is evaluated using the approximate
    linear \ac{PC} model given in \cite[Eq. (4-3)]{imran_energyefficiencyanalysis_2011}
 \begin{equation}\label{eq:pc_model}
    P_c = P_0 + \alpha P
 \end{equation}
where $P_0 = 260$W is the \ac{PC} for zero-load, $\alpha =
2\times4.7$ is the scaling factor term for an antenna with
two-transmission chains and $P$ is the total transmit power when
serving a user with the entire bandwidth.

\begin{table}[!t]
\small
\renewcommand{\arraystretch}{1.3}
\caption{Performance gain using multilevel beamforming for the mass
event scenario} \label{tab:kpisu} \centering
\begin{tabular}{|c|c|c|c|}
    \hline
    m & MUT (Mbps) & CET (Mbps) & PC  (W) \\ \hline
    2 wo. & 7.64 & 1.78 & 397\\\hline
    2 w. & 21.49 (181\%) & 5.52 (210\%) & 334 (-15.92\%)\\\hline
    5 wo. & 7.21 & 1.35 & 400\\\hline
    5 w. & 22.33 (210\%) & 5.59 (312\%) & 331 (-17.24\%)\\\hline
    10 wo. & 6.97 & 1.18 & 402\\\hline
    10 w. & 22.43 (222\%) & 5.31 (349\%) & 331 (-17.51\%)\\\hline
    $+\infty$ wo. & 4.99 & 0.51 & 417 \\\hline
    $+\infty$ w. & 21.85 (337\%) & 4.75 (822\%) & 334 (-19.98\%) \\\hline
\end{tabular}
\end{table}

    The performance results show particularly high gain brought about by multilevel beamforming.
    The \ac{MUT} is improved by a factor varying from 2.81 to 4.38, the \ac{CET} - from 3.1 to 9.22,
    and the \ac{PC} is reduced by 15.9 to 20 percent for m varying from 2
    to $+\infty$ respectively.
    The difference in performance gain between \ac{MUT} and \ac{CET} is due to the fact that
    cell edge users have initially low \ac{SINR} and their \ac{SINR} gain with beam focusing is larger.
    The \ac{PC} is reduced due to the significant reduction in the sojourn time of the users so
    the \ac{BS} transmits less often.

    The performance gains increase with the value of m, namely with
    the importance of the \ac{LoS} component relative to the multipaths' components.
    It is recalled that in an environment rich of scatterers, the initial
    level of beams (i.e. level 0 in Figure \ref{fig:beam_hierarchy}) will benefit
    from higher diversity gain by using an opportunistic scheduler (e.g.
    \ac{PF}) and therefore the gain obtained by the multilevel
    beamforming is smaller. This observation further supports the claim that
    the the multilayer beamforming is of particular interest for
    open type of environment having a significant \ac{LoS}
    propagation.

    \subsection{Rural scenario}

\begin{figure}
        \centering
        \subfloat[Original Cell]{\includegraphics[width=3in]{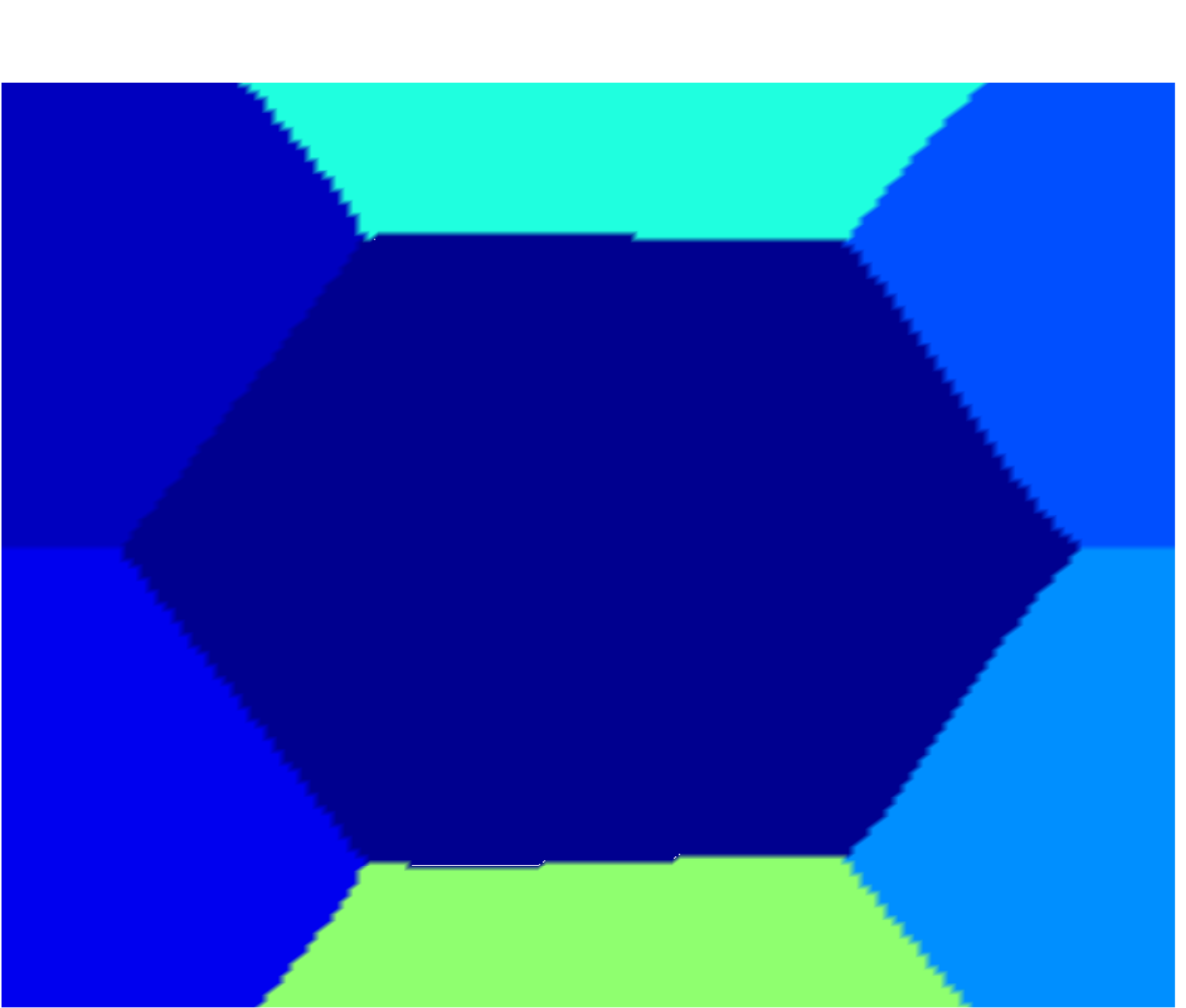}
        \label{fig:beam0r}}
        \hfil
        \subfloat[Level 1 beams ($N_x,N_z$)=(5,14)]{\includegraphics[width=3in]{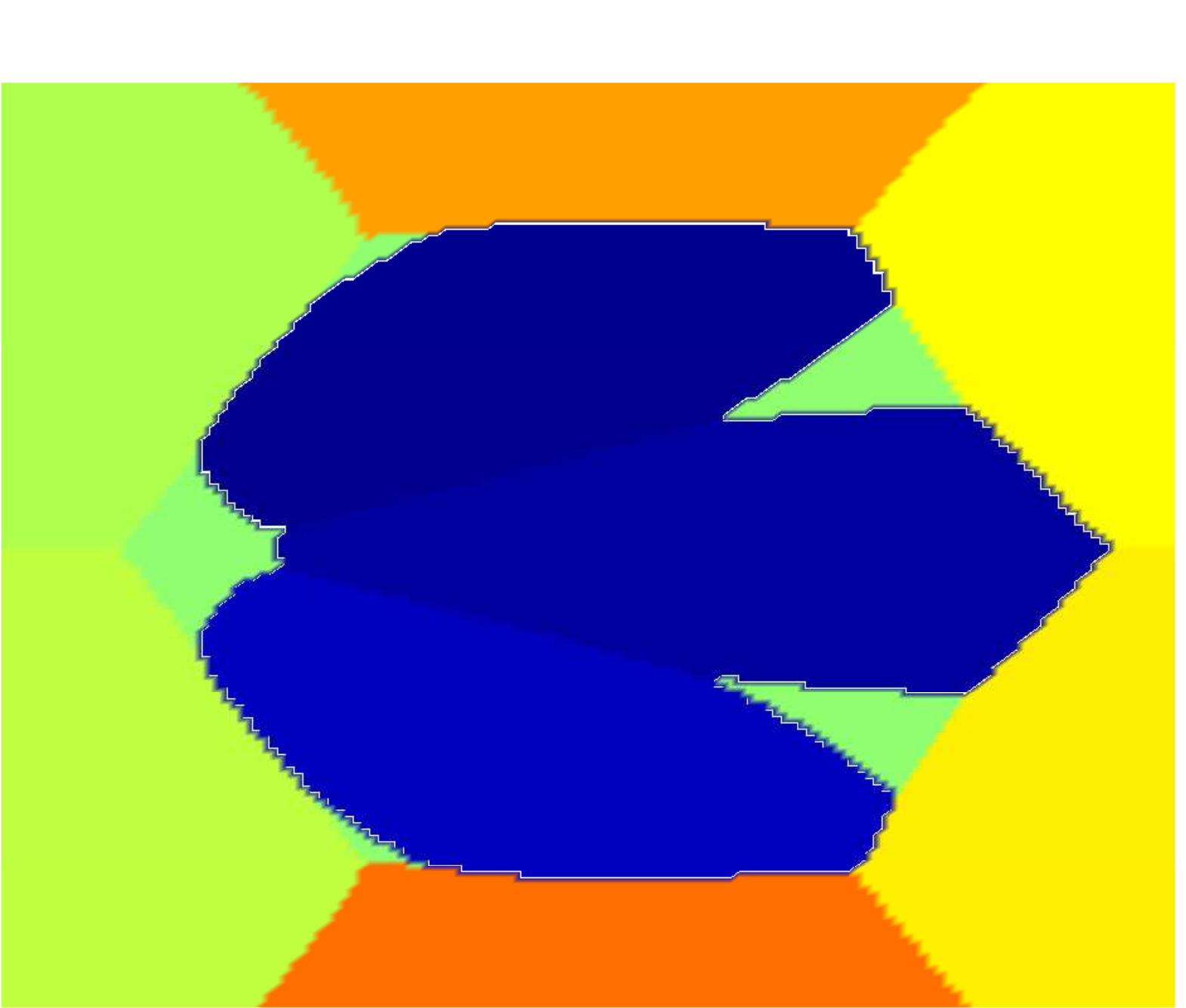}
        \label{fig:beam1r}}
        \hfil
        \subfloat[Level 2 beams ($N_x,N_z$)=(10,14)]{\includegraphics[width=3in]{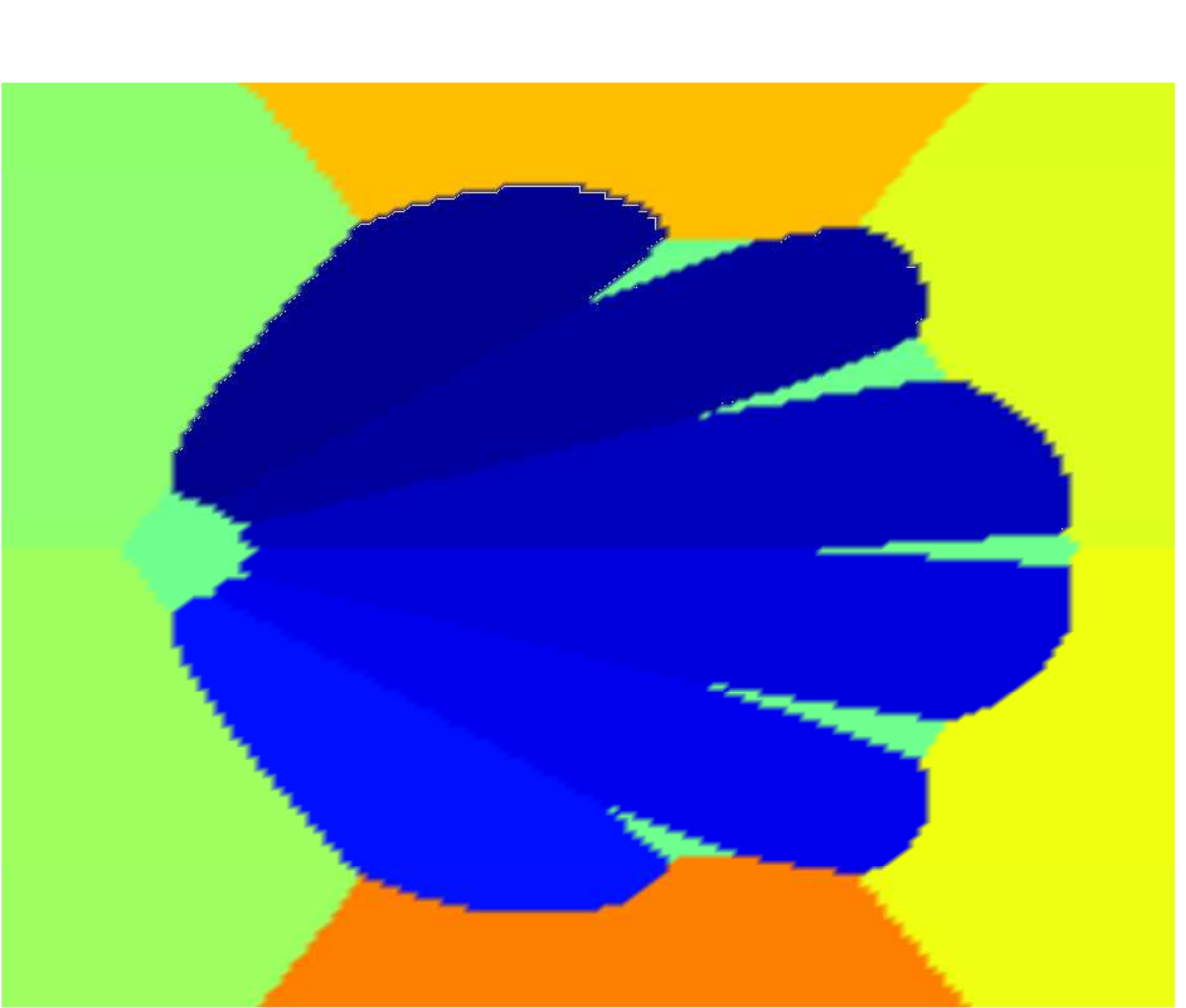}
        \label{fig:beam2r}}
        \hfil
        \subfloat[Level 3 beams ($N_x,N_z$)=(20,14)]{\includegraphics[width=3in]{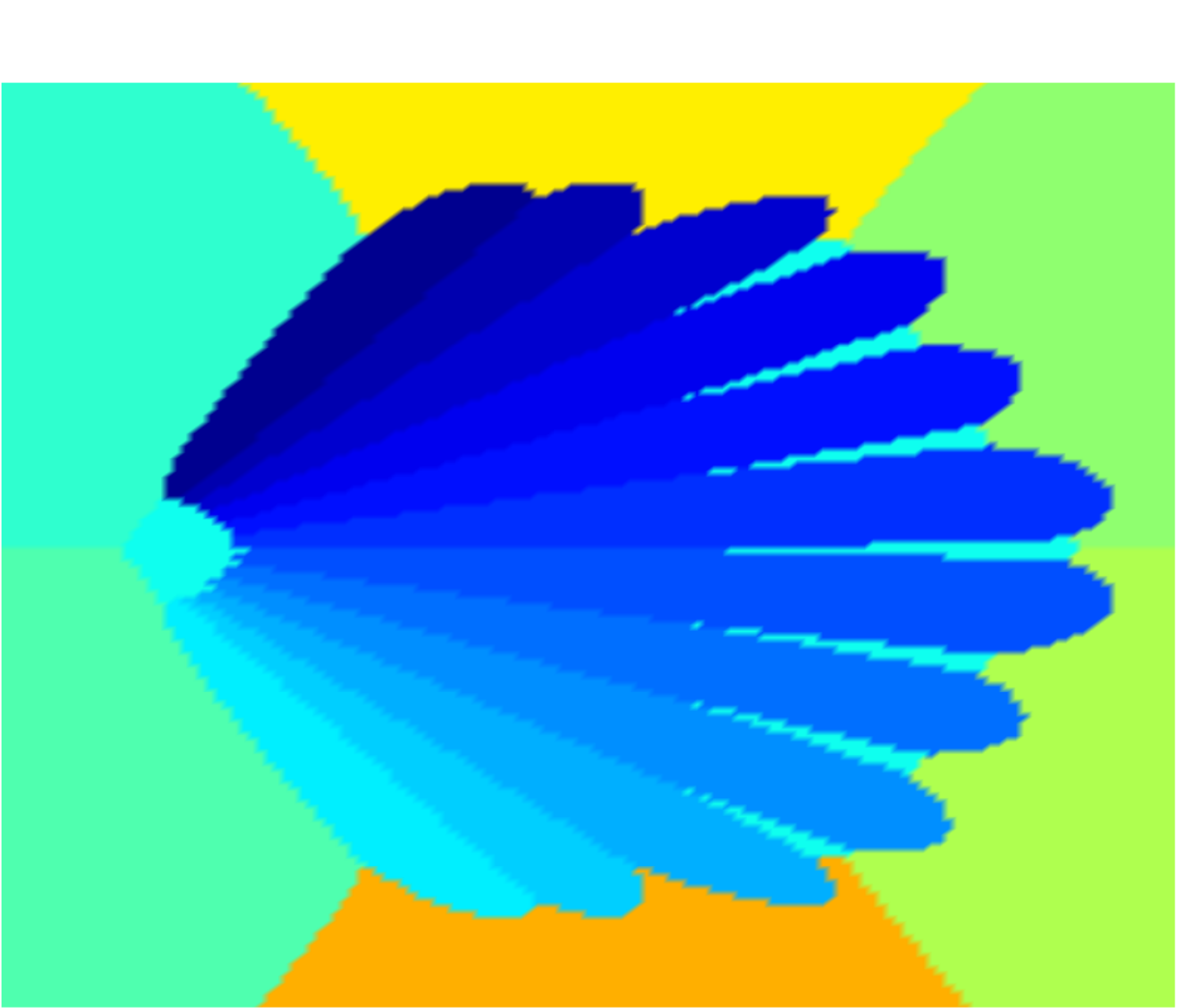}
        \label{fig:beam3r}}
        \caption{Coverage maps for different beamforming levels for the rural scenario}\label{fig:beams_r}
\end{figure}

    The simulation parameters for the rural scenario are summarized in Table \ref{tab:paramsr}.
    Unlike the mass event urban scenario, the bigger dimensions of
    the cell make vertical beam separation complex.
    A modification of the beam direction in elevation by a fraction of
    a degree results in significant difference in its coverage. For
    this reason, we consider multilevel beamforming in the
    horizontal (azimuth) plane, as shown in Figure \ref{fig:beams_r}.

    For the sake of brevity, fast-fading is not considered here. However, similar results as those presented for the mass event urban scenario
    (see Section \ref{sec:urban_perf}) are expected, with performance gains increasing with the shape parameter m of the Nakagami-m fading.

\begin{table}[!t]
\small
\renewcommand{\arraystretch}{1.3}
\caption{Network and traffic characteristics for the rural scenario} \label{tab:paramsr}
\centering
\begin{tabular}{|c|c|}
\hline
Intersite distance & 1732 m \\
\hline
Fast-fading & None \\
\hline
Traffic spatial distribution & uniform \\
\hline
Arrival rate & 2.5 users/s/km$^2$ \\
\hline
\end{tabular}
\end{table}

    Table \ref{tab:kpis} compares performance results for \ac{MUT}, \ac{CET}, and \ac{PC} using different numbers of beamforming levels.
    The performance of level k corresponds to the case where equation \eqref{eq:beam_select} is applied to a highest beam level set to k.
    The performance gain are very high also in the rural scenario.
    For example, for three levels of beams, \ac{MUT} and \ac{CET}
    are increased by 238 and 501 percent. It is noted that the
    gain achieved is lower than that obtained in the mass event urban scenario.
    The reason for this is the smaller number of antenna elements
    used in the rural scenario which results in lower antenna gains. For example, in the third (highest) level,
    the number of antenna elements are $(N_{x,max},N_{z,max})$=(20,14) and $(N_{x,max},N_{z,max})$
    =(12,32) in the rural and the mass event scenarios,
    respectively.

\begin{table}[!t]
\small
\renewcommand{\arraystretch}{1.3}
\caption{Performance gain using multilevel beamforming for the rural
scenario for different beam levels} \label{tab:kpis}
\centering
\begin{tabular}{|c|c|c|c|}
    \cline{2-4}
    \multicolumn{1}{c|}{} & MUT (Mbps) & CET (Mbps) & PC  (W) \\ \hline
    Level 0 & 4.66  & 0.43 & 421 \\ \hline
    Level 1 & 9.9 (112\%)  & 1.15 (168\%) & 388 (-7.93\%) \\ \hline
    Level 2 & 13.23 (184\%) & 1.96 (360\%) & 369 (-12.3\%) \\ \hline
    Level 3 & 15.78 (238\%) & 2.57 (501\%) & 356 (-15.42\%) \\ \hline
\end{tabular}
\end{table}

\section{Conclusion} \label{sec:conclusion}
    A design framework for beam focusing using antenna arrays has been presented in this paper.
    In order to reduce the search complexity among all possible beams when a user is scheduled,
    a multilevel beamforming strategy has been adopted in which the best beam is iteratively selected for each level
    based on the user \ac{CQI}. The multilevel beams' codebook can be constructed offline, as an optimization problem, taking into account
    coverage and interference constraints in each level. A higher level beam covers a fraction (e.g. half)
    of its lower level parent beam. The numerical results show very high performance gains brought about by the multilevel beamforming, both in terms of
    throughput and power consumption. Two scenarios have been evaluated: a mass event urban scenario and a rural scenario.
    The multilevel beamforming solution is well adapted to the
    \ac{FDD} technology, and provides highest gains in environment
    with significant \ac{LoS} component and low level of multipath
    propagation. To achieve highly focused beams, antenna arrays
    with a large number of radiating elements is required. Hence higher
    frequency envisaged in 5G spectrum evolution will make this
    technology particularly attractive.

\bibliographystyle{IEEEtran}
\bibliography{main}

\end{document}